\documentclass[floatfix,twocolumn,aps,prl,reprint,groupedaddress,superscriptaddress]{revtex4-1}
\usepackage{amsmath}
\usepackage{graphicx}

\newcommand{\hh}{\ensuremath{H}}
\newcommand{\kk}{\ensuremath{K}}
\newcommand{\sqh}{\frac{1}{2}}

\begin{document}

\title{Competition between stripe and checkerboard magnetic instabilities in Mn-doped BaFe$_{2}$As$_2$}

\author{G.~S.~Tucker}
\author{D.~K.~Pratt}
\author{M.~G.~Kim}
\author{S.~Ran}
\author{A.~Thaler}
\affiliation{Ames Laboratory and Department of Physics and Astronomy, Iowa State University, Ames, IA, 50011, USA}
\author{G.~E.~Granroth}
\author{K.~Marty}
\affiliation{Oak Ridge National Laboratory, Oak Ridge, TN, 37831, USA}
\author{W.~Tian}
\author{J.~L.~Zarestky}
\affiliation{Ames Laboratory and Department of Physics and Astronomy, Iowa State University, Ames, IA, 50011, USA}
\author{M.~D.~Lumsden}
\affiliation{Oak Ridge National Laboratory, Oak Ridge, TN, 37831, USA}
\author{S.~L.~Bud'ko}
\author{P.~C.~Canfield}
\author{A.~Kreyssig}
\author{A.~I.~Goldman}
\author{R.~J.~McQueeney}
\affiliation{Ames Laboratory and Department of Physics and Astronomy, Iowa State University, Ames, IA, 50011, USA}

\begin{abstract}
Inelastic neutron scattering measurements on $\mathrm{Ba(Fe_{0.925}Mn_{0.075})_{2}As_
{2}}$ manifest spin fluctuations at two different wavevectors in the Fe square lattice, $(\sqh,0)$ and $(\sqh,\sqh)$, corresponding to the expected stripe spin-density wave order and checkerboard antiferromagnetic order, respectively. Below $T_{N}=$80 K, long-range stripe magnetic ordering occurs and sharp spin wave excitations appear at $(\sqh,0)$ while broad and diffusive spin fluctuations remain at $(\sqh,\sqh)$ at all temperatures. Low concentrations of Mn dopants nucleate local moment spin fluctuations at $(\sqh,\sqh)$ that compete with itinerant spin fluctuations at $(\sqh,0)$ and may disrupt the development of superconductivity.
\end{abstract}
\maketitle

Electron and hole doping are common methods used to suppress antiferromagnetism and induce high-temperature superconductivity in copper oxide and iron arsenide compounds. However, the magnetic and chemical disorder introduced by doping can also have a detrimental effect on superconductivity. The iron arsenides give great hope for understanding the phenomenon of high-temperature superconductivity due to their exceptional chemical flexibility.
In BaFe$_{2}$As$_{2}$, antiferromagnetic (AFM) ordering is driven by Fermi surface nesting, resulting in stripe spin-density wave ordering with propagation vector $\mathbf{Q}_{\mbox{\scriptsize stripe}}=(\sqh,0)$ (in Fe square lattice notation).  
A variety of different chemical substitutions can suppress AFM ordering, and the role that the resultant AFM spin fluctuations play in the development of the superconducting state is an open question.

The substitution of Fe with Co acts as an electron donor that detunes the nesting condition,\cite{CLiu10} leading to suppression of the stripe AFM ordering.\cite{Fernandes10} 
The appearance of superconductivity with Co doping is surprising, as doping within the Fe layer may introduce strong disorder effects.\cite{Sefat08,Canfield09}
When Ba$^{2+}$ is substituted by K$^+$, stripe AFM order is suppressed and a superconducting state appears.\cite{Rotter08} Angle-resolved photoemission spectroscopy data on $\mathrm{Ba_{1-x}K_{x}Fe_{2}As_{2}}$ support a picture where K substitutions act as hole donors that detune the Fermi surface pockets connected by $(\sqh,0)$.\cite{Liu08}  Observations of the neutron spin resonance in $\mathrm{Ba(Fe_{1-x}Co_{x})_{2}As_{2}}$ \cite{Lumsden09} and $\mathrm{Ba_{1-x}K_{x}Fe_{2}As_{2}}$ \cite{Christianson08}   
show that superconductivity is coupled to spin fluctuations at the nesting vector $\mathbf{Q}_{\mbox{\scriptsize stripe}}=(\sqh,0)$ for both electron and hole-doped compounds.

Similar to K substitution for Ba,
Mn and Cr substitutions for Fe nominally donate holes and should also detune the Fermi surface nesting.
Intriguingly however, superconductivity never develops; an answer to this mystery may arise from the role that Mn and Cr substitutions play as magnetic impurities in the Fe layer.\cite{Thaler11,Sefat09}   
Neutron diffraction data indicate that Cr doping leads to a (checkerboard) N\'eel antiferromagnetic state with propagation vector $\mathbf{Q}_{\mbox{\scriptsize N\'eel}}=(\sqh,\sqh)$.  In $\mathrm{Ba(Fe_{1-x}Cr_{x})_{2}As_{2}}$, the stripe AFM order gives way to N\'eel AFM order only after a fairly substantial replacement of Fe with Cr, $x>$0.30.\cite{Marty11}   $\mathrm{Ba(Fe_{1-x}Mn_{x})_{2}As_{2}}$ presents a more mysterious case.  While $\mathrm{BaMn_{2}As_{2}}$ is itself a N\'eel antiferromagnet,\cite{Singh09} a different magnetic phase appears at relatively small concentrations of Mn($x>0.10$).\cite{Kim10} The magnetic propagation vector remains the same as that of stripe order, $\mathbf{Q}=(\sqh,0)$, but the tetragonal-orthorhombic structural transition abruptly disappears. This is highly unusual, as the orthorhombic distortion is an expected consequence of magnetoelastic coupling in the presence of stripe AFM order.\cite{Kim10}

In this Letter, inelastic neutron scattering measurements 
examined %were performed to probe 
the effect of Mn substitution on the spin excitations on $\mathrm{Ba(Fe_{1-x}Mn_{x})_{2}As_{2}}$ with $x=0.075$.  
Here we report that Mn doping does not act solely as a hole donor, but instead introduces spin fluctuations at a wavevector $(\sqh,\sqh)$ that is unrelated to the dominant $(\sqh,0)$ Fermi surface nesting of BaFe$_{2}$As$_{2}$.  Spin fluctuations at  $(\sqh,\sqh)$ and $(\sqh,0)$ are observed to coexist, suggesting the Mn dopants act as local magnetic impurities that polarize neighboring Fe/Mn spins.
This %result 
highlights the dual nature of spin fluctuations in unconventional superconductors, which may either promote or disrupt the formation of a superconducting state.  
The introduction of spin fluctuations arising from interactions unrelated to Fermi surface nesting are disruptive for superconductivity.

The sample studied consists of single-crystals of $\mathrm{Ba(Fe_{1-x}Mn_{x})_{2}As_{2}}$ with $x=0.075$ weighing approximately two grams that are co-aligned to within $\sim$1 degree.  
%Previous neutron and x-ray scattering measurements show that t
The sample orders into the stripe AFM structure simultaneously with a tetragonal-orthorhombic transition below $T_{\mbox{\scriptsize N}}=80$\,K,\cite{Kim10} reduced from $135$\,K for the parent $\mathrm{BaFe_{2}As_{2}}$ compound.  
Homogeneity of the Mn doping throughout the sample was verified using wavelength dispersive X-ray spectroscopy, and also by noting the sharpness of the AFM transition.   
Additional details of crystal growth and characterization are described elsewhere.\cite{Thaler11} 
The sample was mounted in a closed-cycle refrigerator for temperature dependent studies on the SEQUOIA, HB1A, and HB3 neutron spectrometers at the Oak Ridge National Laboratory. 
The neutron scattering data are described in the tetragonal $I4/mmm$ coordinate system with 
$\mathbf{Q}=\frac{2\pi}{a}\left(H+K\right)\hat{\imath}+\frac{2\pi}{a}\left(H-K\right)\hat{\jmath}+\frac{2\pi}{c}L\hat{k} = \left(H+K,H-K,L\right)$
where $a=3.97$\,\AA~and $c=12.80$\,\AA~at $T=15$\,K. 
In tetragonal $I4/mmm$ notation, $\mathbf{Q}_{\mbox{\scriptsize stripe}} = \left(\frac{1}{2},\frac{1}{2},1\right)$ [$H$=$\frac{1}{2}$, $K$=$0$] and $\mathbf{Q}_{\mbox{\scriptsize N\'eel}} = \left(1,0,1\right)$ [$H$=$\frac{1}{2}$, $K$=$\frac{1}{2}$]. 
$H$ and $K$ are defined to conveniently describe diagonal cuts in the $I4/mmm$ basal plane as varying $H$ ($K$) corresponds to a longitudinal $\left[H,H\right]$ scan (transverse $\left[K,-K\right]$ scan) through $\mathbf{Q}_{\mbox{\scriptsize stripe}}$.
It can be shown that $H$ and $K$ are the reciprocal lattice units of the Fe square lattice discussed above since
$\mathbf{Q} = \frac{2\pi}{a_{\mathrm{Fe}}}\left[H \frac{1}{\sqrt{2}}\left(\hat{\imath}+\hat{\jmath}\right) + K\frac{1}{\sqrt{2}}\left(\hat{\imath}-\hat{\jmath}\right)\right]=\left(H,K\right)$, where $a_{\mathrm{Fe}}=\frac{a}{\sqrt{2}}$ is the nearest-neighbor Fe-Fe distance.
Thus, $\mathbf{Q}_{\mbox{\scriptsize stripe}} = (\frac{1}{2},0)$ and $\mathbf{Q}_{\mbox{\scriptsize N\'eel}} = (\frac{1}{2},\frac{1}{2})$ in Fe square lattice notation.

\begin{figure}
\includegraphics[width=1.0\linewidth]{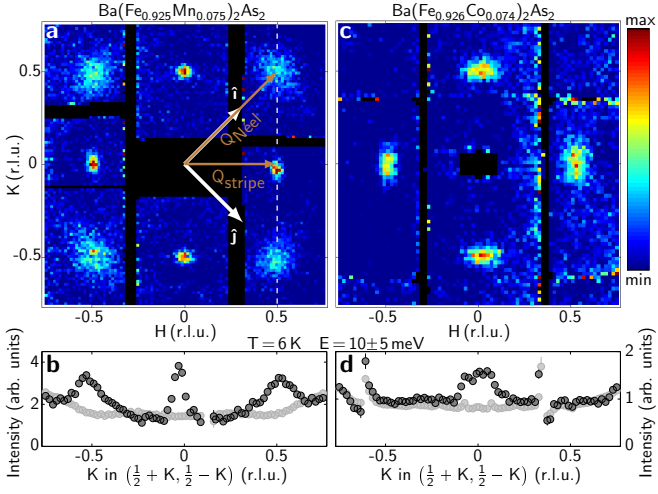}
\caption{\footnotesize 
(a) Spin excitations in $\mathrm{Ba(Fe_{0.925}Mn_{0.075})_{2}As_{2}}$ as measured on SEQUOIA with incident energy $E_{\mathrm{i}}=74.8$\,meV and the crystallographic $\boldsymbol{c}$-axis parallel to the incident beam. 
Data are displayed in the $(H+K,H-K)$ plane and averaged over an energy transfer of $E=$ 5--15\,meV. 
Because of the fixed crystal orientation with incident beam, the $L$ component of the wavevector varies slightly with the in-plane wavevector and energy transfer.
The indicated vectors are
$\mathbf{Q}_{\mbox{\scriptsize stripe}}=(\frac{1}{2},\frac{1}{2},L\!\approx\!1)$ [$H$=$\frac{1}{2}$,$K$=$0$],
$\mathbf{Q}_{\mbox{\scriptsize N\'eel}}=(1,0,L\!\approx\!1)$ [$H$=$\frac{1}{2}$,$K$=$\frac{1}{2}$]
in the coordinate system of the $I4/mmm$ tetragonal lattice.
(b) A cut of the Mn-doped data (black symbols) and estimated phonon background (shaded symbols) along the $\left[\kk,-\kk\right]$-direction through $\left(\frac{1}{2},\frac{1}{2}\right)$, as indicated by the dashed line in (a). 
(c) Spin excitations in $\mathrm{Ba(Fe_{0.926}Co_{0.074})_{2}As_{2}}$ averaged over an energy transfer range of 5--15\,meV as measured on the ARCS spectrometer with $E_\mathrm{i}=49.8$\,meV and the $c$-axis parallel to the incident beam.  
(d) A cut of the Co-doped data (black symbols) and background (shaded symbols) along the $\left[\kk,-\kk\right]$-direction through $\left(\frac{1}{2},\frac{1}{2}\right)$.  
In (a) and (c), an estimate of the instrumental background has been subtracted.}
\label{fig1}
\end{figure}

Figure 1(a) shows background subtracted data in the stripe AFM ordered phase at $T=6$\,K and an energy transfer $E=10$~meV.  Sharply defined peaks are observed near $\mathbf{Q}_{\mbox{\scriptsize stripe}} =(\frac{1}{2},\frac{1}{2},1)$ and symmetry related wavevectors corresponding to spin waves in the stripe AFM state 
.%[with wavevector $\mathbf{Q}_{\mbox{\scriptsize stripe}}=(\sqh,0)$ in the Fe square lattice notation].  
Broad inelastic scattering is also observed at $\mathbf{Q}_{\mbox{\scriptsize N\'eel}} =(1,0,1)$ and symmetry related wavevectors 
.%[with wavevector $\mathbf{Q}_{\mbox{\scriptsize N\'eel}}=(\sqh,\sqh)$ in Fe square lattice notation].  
A transverse cut along the $[\kk,-\kk]$-direction in Fig.~1(b) highlights the strong inelastic scattering at both $\mathbf{Q}_{\mbox{\scriptsize stripe}}$ and $\mathbf{Q}_{\mbox{\scriptsize N\'eel}}$.  Previously published data on Co-doped $\mathrm{Ba(Fe_{0.926}Co_{0.074})_{2}As_{2}}$ find no such excitations at $\mathbf{Q}_{\mbox{\scriptsize N\'eel}}$.\cite{Li10}  At this comparable doping level, the Co-doped sample is paramagnetic and superconducting, and Figs. 1(c) and 1(d) show only a broad peak at $\mathbf{Q}_{\mbox{\scriptsize stripe}}$ originating from stripe spin fluctuations common to all iron-based superconductors.\cite{Li10}  

It is at first unclear whether the inelastic scattering intensity observed in the Mn-doped sample at $\mathbf{Q}_{\mbox{\scriptsize N\'eel}} $ is magnetic or nuclear in origin, since $(1,0,1)$ 
is a nuclear Bragg peak in the $I4/mmm$ space group. The absence of any inelastic peak in Co-doped materials near $\mathbf{Q}_{\mbox{\scriptsize N\'eel}} $ suggests that it does not arise from phonons in the vicinity of the $(1,0,1)$ nuclear peak.  
Perhaps the strongest evidence for the magnetic origin of this feature comes from measurements 
%of the intensity 
along the $(1,0,L)$ direction using the HB1A triple-axis spectrometer, shown in Fig.~2(a).  At $T=10$\,K and $E=3$\,meV, the intensity manifests a sinusoidal modulation peaked at odd values of $L$, as expected for antiferromagnetic spin correlations between the layers.  The decay of the signal at large $L$ follows the Fe$^{2+}$ magnetic form factor, thereby confirming their magnetic nature.

\begin{figure}
\includegraphics[width=0.625\linewidth]{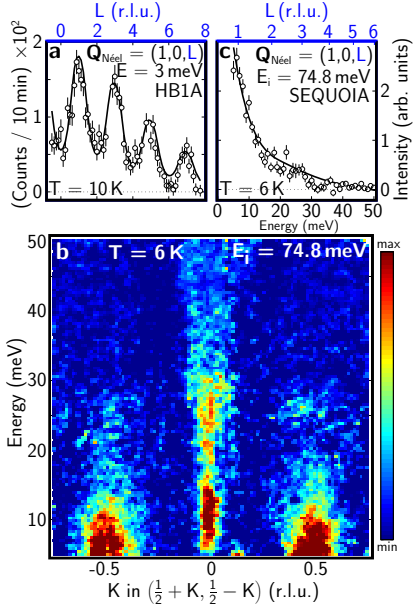}
\caption{\footnotesize
(a) Spin excitations in $\mathrm{Ba(Fe_{0.925}Mn_{0.075})_{2}As_{2}}$ measured along $[1,0,L]$ at $T=10$ K  on HB1A  with $E=3$\,meV, $E_\mathrm{i}=14.7$\,meV, $48'$-$40'$-$80'$-$136'$ horizontal collimation, and the sample mounted in the $[100]$-$[001]$ scattering plane. (b) Spectrum of spin excitations emanating from $\mathbf{Q}_{\mbox{\scriptsize N\'eel}}$ and $\mathbf{Q}_{\mbox{\scriptsize stripe}}$ as measured on SEQUOIA after averaging over the range $\hh= 0.50\pm0.05$.
(c) Energy spectrum at $\mathbf{Q}_{\mbox{\scriptsize N\'eel}}$ from SEQUOIA data after averaging over the Q-region $\kk=-0.50\pm0.05$, $\hh=0.50\pm0.05$.
The lines in (a) and (c) are a fit to a model of paramagnetic spin fluctuations as described in the text.
}
\label{fig2}
\end{figure}

Fig.~2(b) shows that the magnetic spectrum at $\mathbf{Q}_{\mbox{\scriptsize stripe}}$ consists of steep spin waves associated with the long-range stripe AFM order whereas, at $\mathbf{Q}_{\mbox{\scriptsize N\'eel}}$, figs.~2(b) and (c) indicate that the spectrum has a quasielastic or relaxational form.  Therefore, Mn-doping introduces short-ranged checkerboard-like spin correlations with wavevector $\mathbf{Q}_{\mbox{\scriptsize N\'eel}}$ that are purely dynamic and coexist with the  long-range stripe AFM order.

To 
clarify %better investigate 
the relationship between magnetism at $\mathbf{Q}_{\mbox{\scriptsize N\'eel}}$ and $\mathbf{Q}_{\mbox{\scriptsize stripe}}$, we studied the temperature dependence of the spin fluctuations, as illustrated in Fig.~3(a)-(e).  Magnetic fluctuations at $\mathbf{Q}_{\mbox{\scriptsize N\'eel}}$ are weakly temperature dependent and persist up to at least 300\,K.  As expected, the excitations at $\mathbf{Q}_{\mbox{\scriptsize stripe}}$ become broader above $T_{\mbox{\scriptsize N}}=80$\,K and paramagnetic stripe spin fluctuations become nearly washed out at 300\,K.  This broadening occurs more strongly along the $[\kk,-\kk]$-direction transverse to $\mathbf{Q}_{\mbox{\scriptsize stripe}}$, signaling a significant temperature dependence of the in-plane anisotropy of the stripe spin fluctuations.

\begin{figure}
\includegraphics[width=0.625\linewidth]{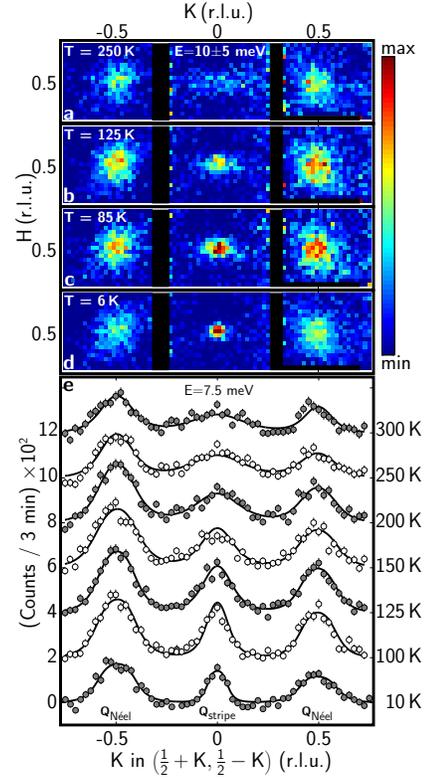}
\caption{\footnotesize
Spin excitations in $\mathrm{Ba(Fe_{0.925}Mn_{0.075})_{2}As_{2}}$ measured on SEQUOIA at (a) 250\,K, (b) 125\,K, (c) 85\,K, and (d) 6\,K in a section of the $(H+K,H-K)$ plane after averaging over an energy transfer range of $10\pm5$\,meV, and (e) measured on HB3 along the $[\kk,-\kk]$-direction through $(\frac{1}{2},\frac{1}{2},1)$ at several different temperatures (offset) and an energy transfer $E=7.5$\,meV with $E_\mathrm{f}=30.5$\,meV, collimation $48'$-$60'$-$80'$-$120'$, and the sample mounted in the [112]-[1\={1}0] scattering plane.  Lines in (e) are fits to a model of paramagnetic spin fluctuations as described in the text.}
\label{fig3}
\end{figure}

The temperature dependence of the imaginary part of the dynamic magnetic susceptibility $\chi''(\mathbf{Q}_{\mbox{\scriptsize N\'eel}},E)$ and $\chi''(\mathbf{Q}_{\mbox{\scriptsize stripe}},E)$ is  shown in Figs.~4(a) and 4(b), respectively.   The susceptibility is obtained from neutron scattering data according to the formula 
\begin{equation}
S(\mathbf{Q},E) \propto f^{2}(Q)\chi''(\mathbf{Q},E)(1-e^{-E/kT})^{-1}
\label{eq_S}
\end{equation}
where $S(\mathbf{Q},E)=I(\mathbf{Q},E)-B(\mathbf{Q},E)$ is the magnetic intensity, $I(\mathbf{Q},E)$ is the measured intensity, $B(\mathbf{Q},E)$ is a background function estimated from scans performed well away from the magnetic scattering features, and $f(Q)$ is the magnetic form factor of an $\mathrm{Fe}^{2+}$ ion.\cite{Brown}  The susceptibility at both wavevectors grows upon cooling, indicating competing magnetic instabilities.  
%At lower temperatures, t
The stripe AFM correlations become dominant and the susceptibility reaches a maximum consistent with critical scattering near the onset of stripe magnetic order at $T_{\mbox{\scriptsize N}}$. Surprisingly, the stripe AFM ordering has little or no effect on the susceptibility at $\mathbf{Q}_{\mbox{\scriptsize N\'eel}}$, which continues to grow as the temperature is lowered.  In other systems with competing magnetic instabilities, such as the ferromagnetic and antiferromagnetic interactions in colossal magnetoresistive manganites, the ordering of the dominant instability quickly suppresses competing fluctuations.\cite{Kajimoto98} 

\begin{figure}
\begin{center}
\includegraphics[width=0.85\linewidth]{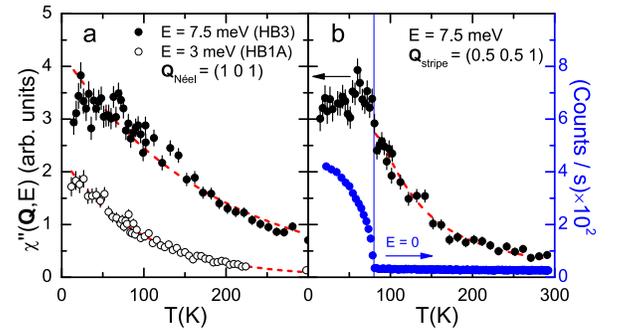}
\caption{\footnotesize (a) $\chi''(\mathbf{Q}_{\mbox{\scriptsize N\'eel}},E)$ measured at $E=3$\,meV on HB1A (open symbols) and 7.5\,meV on HB3 (closed symbols) as a function of temperature.
(b) Temperature dependence of $\chi''(\mathbf{Q}_{\mbox{\scriptsize stripe}},E=7.5$\,meV) (black circle symbols) as measured on HB3.   The intensity of the magnetic Bragg peak at $\mathbf{Q}_{\mbox{\scriptsize stripe}} =(\frac{1}{2},\frac{1}{2},1)$ (blue square symbols) indicates the development of stripe magnetic order below $T_{\mbox{\scriptsize N}}$. The dashed lines in (a) and (b) are guides to the eye.
}
\end{center}
\label{fig4}
\end{figure}

%Details of t
The length and energy scales of the paramagnetic spin fluctuations is obtained by fitting both the \mbox{SEQUOIA} and \mbox{HB3} data measured above $T_{\mbox{\scriptsize N}}$ to a susceptibility function that has been successfully used to describe iron arsenide compounds.\cite{Inosov09,Diallo10} 
The susceptibility of the Fe layers near $\mathbf{Q}_{\mbox{\scriptsize N\'eel}}$ and $\mathbf{Q}_{\mbox{\scriptsize stripe}}$ is described by
\begin{equation}
\chi''(\mathbf{Q}_{\mbox{\scriptsize N\'eel}}+\mathbf{q},E)=\frac{A_{\mbox{\scriptsize N}}E\Gamma_{\mbox{\scriptsize N}}}{E^2+\Gamma_{\mbox{\scriptsize N}}^2\left[1+\xi_{\mbox{\scriptsize N}}^2 q^2\right]^2}
\label{eq_xi2}
\end{equation}
\begin{equation}
\chi''(\mathbf{Q}_{\mbox{\scriptsize stripe}}+\mathbf{q},E)=\frac{A_{\mbox{\scriptsize S}}E\Gamma_{\mbox{\scriptsize S}}}{E^2+\Gamma_{\mbox{\scriptsize S}}^2\left[1+\xi_{\mbox{\scriptsize S}}^2(q^2+\eta q_x q_y)\right]^2}
\label{eq_xi1}
\end{equation}
where $\xi_{i}$, $\Gamma_{i}$, and $A_{i}$ ($i=\mbox{S}$ or $\mbox{N}$) are the spin-spin correlation length, relaxation rate, and scale factor, respectively.  The parameter $\eta$ characterizes the two-fold anisotropy of the correlation length expected at $\mathbf{Q}_{\mbox{\scriptsize stripe}}$.  The in-plane momentum transfer vector, $\mathbf{q}=(q_x,q_y)$, is defined relative to either $\mathbf{Q}_{\mbox{\scriptsize N\'eel}}$ or $\mathbf{Q}_{\mbox{\scriptsize stripe}}$.

\begin{figure}
\begin{center}
\includegraphics[width=0.85\linewidth]{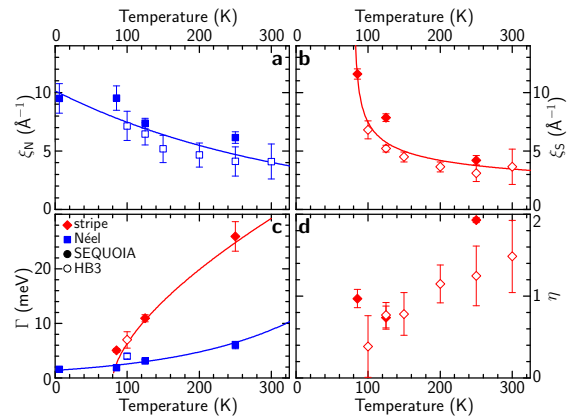}
\caption{Temperature dependence of the model parameters (a) $\xi_\mathrm{N}(T)$ (b) $\xi_\mathrm{S}(T)$, (c) $\Gamma_{i}(T)$, (d) $\eta$ obtained by fitting the inelastic neutron scattering data. Blue square symbols correspond to parameters for $\chi''_\mathrm{N}$ and red diamond symbols for $\chi''_\mathrm{S}$. Closed symbols were determined from \mbox{SEQUOIA} data and open symbols from HB3 data.  Lines in (a) and (b) correspond to fits to Eqns.~(\ref{eq_xiNT}) and (\ref{eq_xiT}), respectively.  The lines in (c) correspond to those obtained after applying the scaling relation Eqn.~(\ref{eq_Gam}).
}
\label{fig5}
\end{center}
\end{figure}

In the theory for itinerant antiferromagnets, the relaxation width is determined by the scaling relation 
\begin{equation}
\Gamma_{i}(T) = \frac{\gamma_{i} a^2}{\xi_{i}(T)^{2}} 
\label{eq_Gam}
\end{equation}
where $\gamma_{i}$ is a characteristic spin fluctuation energy.\cite{Moriya}  For stripe spin fluctuations above the ordering temperature $T_{\mbox{\scriptsize N}}$, the correlation length is expected to obey the form for critical scattering
\begin{equation}
\xi_{\mbox{\scriptsize S}}(T) = \xi_{\mbox{\scriptsize S}0} \left(\frac{T-T_{\mbox{\scriptsize{N}}}}{T_{\mbox{\scriptsize{N}}}}\right)^{-\nu}
\label{eq_xiT}
\end{equation}
where $\nu$ is the critical exponent.  Since there is no ordering temperature associated with spin fluctuations at $\mathbf{Q}_{\mbox{\scriptsize N\'eel}}$, we assume that its correlation length follows an empirical exponentially decaying form:
\begin{equation}
\xi_{\mbox{\scriptsize N}}(T)=\xi_{\mbox{\scriptsize N}0} e^{-T/T^*}.
\label{eq_xiNT}
\end{equation}
\indent
The temperature dependence of model parameters obtained by fitting the SEQUOIA and HB3  data to Eqns.~(\ref{eq_S}--\ref{eq_xi1}) are shown in Fig.~\ref{fig5}, representative fits are shown as solid lines in Fig.~\ref{fig3}(e).
Solid lines in Fig.~\ref{fig2} are fits which also include $L$-dependence of the susceptibility as described in Ref.~\onlinecite{Diallo10}.
Subsequent fits to $\xi_{\mbox{\scriptsize S}}(T)$ and $\xi_{\mbox{\scriptsize N}}(T)$ using Eqn.~(\ref{eq_xiT}) and Eqn.~(\ref{eq_xiNT}), respectively, allowed us to verify the scaling relation for $\Gamma_{i}(T)$ in Eqn.~(\ref{eq_Gam}). We obtained the following temperature independent parameters; $\nu=0.32\pm0.05$, $\xi_{\mbox{\scriptsize S}0}=4.8\pm0.4$ \AA, $\xi_{\mbox{\scriptsize N}0}=10\pm1$ \AA, $T^*=330\pm80$ K, $\gamma_{S}=23\pm3$ meV, $\gamma_{N}=8.3\pm0.5$ meV.

Based on the fitting results in Fig.~5(a), we see that the spin fluctuations at $\mathbf{Q}_{\mbox{\scriptsize N\'eel}}$ exist in nanoscale regions with a characteristic size of  $\sim$ 10~\AA~at 10~K.    Mn doping therefore nucleates small fluctuating regions characterized by dominant nearest-neighbor AFM (or G-type) exchange interactions.  This behavior is consistent with $\mathrm{BaMn_{2}As_{2}}$ itself, which is a G-type AFM insulator with large local Mn$^{2+}$ moments and high $T_{\mbox{\scriptsize N}}=625$\,K.\cite{Singh09} Thus, rather than acting solely as a hole donor, Mn dopants act as magnetic impurities that couple to neighboring spins by magnetic interactions that are unrelated to the Fermi surface nesting at 
$\mathbf{Q}_{\mathrm{stripe}}$. 

The positive value of $\eta$ in Fig.~5(d) indicates that stripe spin fluctuations have a shorter correlation length in the ferromagnetic bond direction, transverse to $\mathbf{Q}_{\mbox{\scriptsize stripe}}$.  Electronic structure calculations of the generalized susceptibility predict that $\eta<0$ in hole-doped compounds, i.e. that the correlation length is shorter along the antiferromagnetic (longitudinal) direction \cite{Park10}.  This has recently been observed in (Ba$_{1-x}$K$_x$)Fe$_2$As$_2$\cite{Zhang11}, but is clearly not the case here and provides further evidence that Mn is not acting as a hole donor.  Values of $\eta$ approach the critical ratio for the transformation from stripe to N\'eel ordering in a local moment model ($\eta=J_{1}/J_{2}=2$) at high temperatures and signify a competition between stripe and N\'eel ordering.  The suppression of $T_{\mbox{\scriptsize N}}$ with Mn doping,\cite{Kim10} while qualitatively similar to Co doping,\cite{Fernandes10} is more likely a consequence of these competing magnetic interactions rather than detuning of the Fermi surface nesting.  It is interesting to ponder whether the presence of 
spin fluctuations at $\mathbf{Q}_{\mbox{\scriptsize N\'eel}}$ would prevent the appearance of superconductivity in the Mn-doped system if $T_N$ could be suppressed further.  However, slightly higher Mn doping ($x=0.10$) instead leads to the development of a more stable ordered magnetic phase.

The case of Mn-doped BaFe$_2$As$_2$ has some interesting parallels to the $\mathrm{Fe_{1+y}Te_{1-x}Se_{x}}$ system.  $\mathrm{Fe_{1+y}Te_{0.6}Se_{0.4}}$ is a superconductor characterized by spin fluctuations near $\mathbf{Q}_{\mbox{\scriptsize stripe}} = (\sqh,0)$ (in Fe square lattice notation).\cite{Liu10}  As the relative Se content is reduced towards $\mathrm{Fe_{1+y}Te}$, superconductivity is suppressed in favor of large moment magnetic ordering with Fe square lattice wavevector near $(\frac{1}{4},\frac{1}{4})$ that is unrelated to Fermi surface nesting.\cite{Xia09} Intermediate compositions of $\mathrm{Fe_{1+y}Te_{1-x}Se_{x}}$ are characterized by spin fluctuations at both wavevectors.\cite{Lumsden10,Xu10}  Therefore, both Fe(Te,Se) and $\mathrm{BaFe_{2}As_{2}}$ systems allow superconductivity to be induced by the appropriate chemical substitution, but other chemical substitutions may result in the appearance of local moment magnetism that is deleterious for superconductivity. 

\begin{acknowledgments}
The work at Ames Laboratory was supported by the U.S. Department of Energy, Office of Basic Energy Science, Division of Materials Sciences and Engineering under Contract No. DE-AC02-07CH11358. Work at Oak Ridge National Laboratory is supported by U.S. Department of Energy, Office of Basic Energy Sciences, Scientific User Facilities Division.
\end{acknowledgments}

\end{document}